# Heatpipe-cooled in-vacuum electromagnet for quantum science experiments

Kenneth Nakasone (中曽根ケネス)[1], Paola Luna[1], Andrei Zhukov[1], Matthew Tao[2], Garrett Louie[2], Cristian D. Panda[1a)]

Quantum inertial sensors test general relativity, measure fundamental constants, and probe dark matter and dark energy in the laboratory with outstanding accuracy. Their precision relies heavily on carefully choreographed quantum control of the atomic states with a collection of lasers, microwaves, electric and magnetic fields. Making this technology available outside of the laboratory would unlock many applications, such as geophysics, geodesy and inertial navigation. However, this requires an apparatus of reduced size, weight, power use and increased robustness, modularity and ease-of-use. Here, we describe the design and implementation of an in-vacuum electromagnet able to create the magnetic fields necessary for various quantum control operations, such as magneto-optical trapping or magnetic levitation to assist evaporative cooling. Placing the electromagnet inside the vacuum chamber has significant advantages, such as fast switching times that are not limited by induced current inside the vacuum chamber metal, reduced size, weight and power usage. However, dissipating the heat produced typically requires complex designs that include bulky metal heatsinks or cooling using water or cryogens. Our design implements heatpipes in a compact, low-vibration and robust apparatus, which use a phase transition in the working fluid to achieve thermal conductivity that is more than one hundred times larger than that of typical bulk metal. We show that the setup can conduct more than 50 W of thermal power in a configuration that provides ample optical access and is compatible with the ultra-high vacuum requirements of atomic and molecular experiments.

[1] Wyant College of Optical Sciences, University of Arizona, Tucson, 85721, AZ, USA
[2] Department of Physics, University of California, Berkeley, 94720, CA, USA

[a)] Author to whom correspondence should be addressed: cpanda@arizona.edu.

# Introduction

Atomic and molecular experiments often use complex quantum control steps that involve several optical, magnetic and electric fields. Precisely controlled magnetic fields are therefore ubiquitous in these experimental setups and have been used e.g. for magnetic trapping,[1,2] magneto-optical trapping (MOT)[3,4] and evaporative cooling.[5,6] In particular, atom interferometers commonly use magnetic field offsets of a few Gauss to maintain a spin quantization axis during interferometry, and gradients of 10-20 G/cm to gather atoms inside magneto-optical traps.[7–9] In addition, many experiments use electromagnets for evaporative cooling to condense atoms into a Bose-Einstein condensate.[10,11]

Atom fountains have demonstrated measurement times of a few seconds, limited by the available free-fall time of atoms inside meter-tall vacuum chambers. Alternatively, interferometry with atoms held by an optical lattice[12–16] can now measure quantum spatial superposition states with coherence at the one-minute timescale. Recent studies[7] have identified key parameters that limit the coherence in lattice atom interferometers, including thermal atom motion inside the optical trap and oscillatory tilts of the optical lattice. Higher coherence may be achieved by reducing the atom sample temperature below the recoil limit using evaporative cooling[17,18] and by implementing lattice vibration stabilization methods to reduce the effects of tilts.[19] We will perform evaporative cooling in our lattice atom interferometer following a procedure that combines magnetic levitation using a 31 G/cm magnetic field gradient with an optical dipole trap.[18]

In this manuscript, we describe the design and experimental implementation of a heatpipe-cooled in-vacuum electromagnet that will produce the necessary magnetic fields for the magneto optical trap (MOT) and for magnetic levitation during evaporation. The electromagnet is designed to fit inside a commercial spherical octagon steel vacuum chamber, in a compact geometry that is compatible with optical access for several laser beams used for various atom sample preparation and detection steps. The electromagnet uses as little as 50 W of power to generate field offsets and gradients as high as 138 G and 38 G/cm respectively. The coil achieves sub-ms turn-on times by deliberately breaking current loops (Eddy currents) with insulating materials. The thermal design uses heatpipes to quickly dissipate heat, reducing excess heating and material outgassing. Our setup achieves operating pressures of order $10^{-10}$ torr, sufficient for quantum science experiments with minute-long atomic lifetimes.

# Design and Implementation

We identify a series of constraints and parameters that inform the choices and avenues we took during the design and implementation process. Our lattice atom interferometer uses an optical resonator to provide the uniform optical lattice potential[7,20] needed to hold atoms in a coherent quantum spatial superposition state. Thus, the resonator mirrors need to be mounted on a stable, rigid structure, which also serves the role of a vacuum chamber.

We use a commercial Kimball 8” Extended Spherical Octagon vacuum chamber (Model MCF800-ExtOct-G2C8A16), which consists of two primary 8” ConFlat™ (CF) ports surrounded by secondary CF ports in an octagonal layout. The origin of our coordinate system is defined as the geometric center of the Kimball chamber. The coordinate system axes are shown in Figure 1a and 1c, where the *z*-axis is aligned against gravity, along the centers of the vertical 4.5” ports, the

*x*-axis is horizontal along the centers of the 8” ports, the *y*-axis is along the centers of the horizontal 4.5” ports.

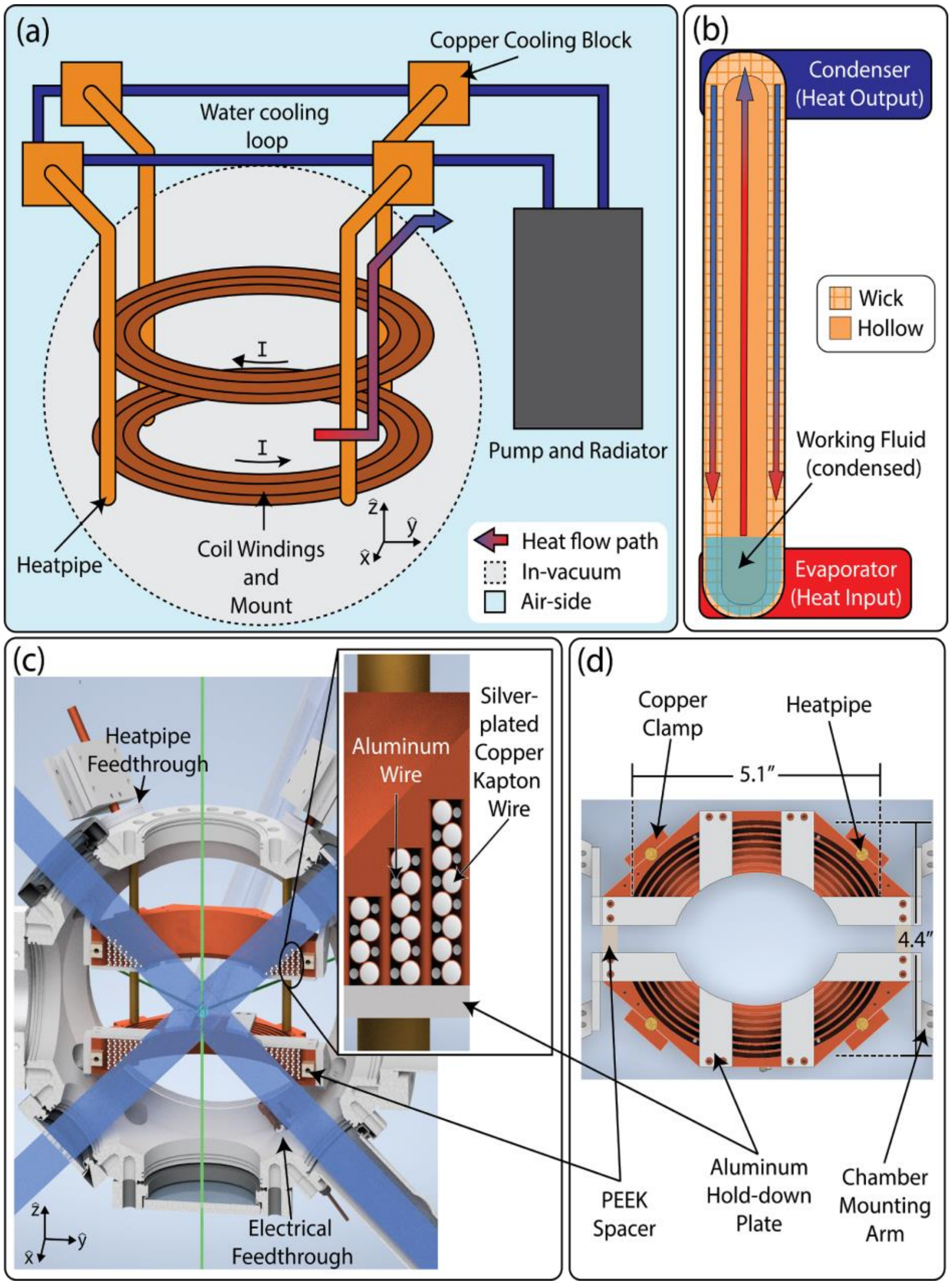

**Fig 1. Experiment Design. (a) Schematic of electromagnet apparatus.** Elliptical coils with current flowing in opposing directions are installed inside a vacuum chamber. Heatpipes are used to carry the dissipated heat outside of vacuum, where it is transferred out of the system by a water cooling loop. **(b) Heatpipe working principle.** The heatpipe efficiently moves heat through evaporation and condensation of the working fluid. **(c) Cross-sectional render of the electromagnet.** Kapton coated wires are wound in channels machined in a copper mounting block housed in a vacuum chamber and carry electrical current supplied through electrical feedthroughs to create the magnetic field offsets and gradients necessary in the atomic experiment. The electromagnet is rigidly mounted to the vacuum chamber and provides ample optical access for the laser beams used for quantum control (blue, green). Spacers made out of polyether ether ketone (PEEK) break electrically conductive loop (Eddy) currents around the copper mounting blocks, allowing for fast switching. The atom sample (cyan sphere) is located at the center of the electromagnet. **(d) Top view of electromagnet mount**, highlighting the elliptical shape.

Placing the electromagnet outside of the vacuum chamber has two disadvantages: (a) the distance from the atoms to the coils is much larger so that achieving the necessary magnetic field gradients requires relatively high powers of several kW and (b) the switch on/off time of the magnet is typically limited to a few ms by the induced current loops (Eddy current) in the conductive vacuum chamber. Both of these can be addressed by high-current coil driver circuits with advanced features that allow for the cancellation of Eddy currents.[21] To avoid these challenges, we install our in-vacuum electromagnet inside the vacuum chamber (Figure 1).

We create a near-quadrupole magnetic field at the center of the chamber with a pair of coils placed equidistant from the atom source. The coil shape and thickness is restricted by the optical access (see Appendix A) required for laser beams used in various atom sample preparation steps.[7,20] MOT laser beams (1" diameter) travel along the horizontal and diagonal axes of the octagon chamber (blue in Figure 1c). Vertical laser beams resonant inside an optical cavity are used to enact atomic Raman beamsplitter operations. Laser beams aligned diagonally in the horizontal plane create the potentials necessary for Raman Sideband Cooling and evaporative cooling. Imaging is performed using a large numerical aperture lens (2" diameter) placed in the horizontal plane. These beams carve out a significant amount of optical access, limiting the geometry of the electromagnet.

FItting the coils inside the remaining vacuum chamber space yields an elliptical geometry with the major axis along $y$ and minor axis along $x$. The windings are stacked in a right triangle cross section (see Figure 1c) that maximizes the cross sectional coil area. Each coil consists of 43 windings of Kapton-insulated silver-plated stranded copper wire (Accu-Glass #100715, 14 AWG), chosen for vacuum compatibility. Each coil of the electromagnet is connected to a 1.33" electrical feedthrough (EFT1223092) via beryllium copper (BeCu) push connectors (Lesker FTASSC050). On the air-side, each coil is connected to a commercial current supply (Agilent 3633).

The kapton wire is tightly wound in channels machined in elliptical copper mounting blocks to obtain good thermal conductivity.[22,23] We use OFE copper (i.e. copper 101) for its vacuum compatibility and high bulk thermal conductivity. To improve heat transfer from the kapton wire to the copper mounting blocks, we insert aluminum wire (0.02" diameter) in the gaps between the kapton wire and mounting block channels. We alternate between filling the inner and outer gap every winding layer (illustrated in Figure 1c) to improve vertical packing density. We further secure the windings by applying pressure with aluminum hold-down plates pulled down by aluminum wire (Figure 1d).

When the copper mounting block for a coil is a single copper piece, Eddy currents limit switching time to a few tens of ms. We break these current loops by splitting the mounting block into two halves, joined by an insulating polyether ether ketone (PEEK) spacer (Figure 1c and 1d). We secure the copper mounting blocks to the vacuum chamber using two-piece arm assemblies with 0.18" deep 4-40 holes tapped into 1.33" conflat blanks. The blanks are attached to the chamber on the 1.33" ports adjacent to the horizontal 4.5" ports (Appendix B).

We use heatpipes to passively transfer heat from the mounting blocks to the outside of the vacuum chamber. A heatpipe is a sealed tube that efficiently transfers heat using a working fluid and capillary action (see Figure 1b for working principle). When heat is applied to one end (the evaporator), the fluid inside absorbs it and vaporizes, carrying thermal energy to the cooler end (the condenser), where it condenses and releases heat. The liquid is then drawn back to the hot end through a wick structure via capillary action, enabling a continuous, passive cycle.[24,25]

The heatpipe thermal conductivity can be as much as three orders of magnitude higher than metal bulk conduction or radiation. In our case, we measure thermal conductivities of up to 14 kW/(m·K) (see section Thermal and Vacuum Testing and Appendix E below), a large improvement over solid copper (0.39 W/m·K) or solid aluminum (0.2 W/m·K). They provide reduced vibrations, reduced complexity and minimal risk of failure when compared to similar water or cryogenic flow cooling methods. Aligning the heatpipes with gravity maximizes their convective cooling efficiency.

Each of the 4 copper heatpipes is passed through to the center of a 1.33" CF blank and brazed to create a vacuum-tight seal, forming a feedthrough assembly (custom manufactured by Noren Thermal Solutions). They are 6 mm diameter, 2.8" long on the air-side and 6" long on the vacuum side, filled with methanol. Using methanol as heatpipe fluid improves thermal conductivity below ambient (20$^{o}$ C) due to its lower freezing point than more commonly used water. The vacuum side is bent such that, when installed in the chamber, each heatpipe is oriented vertically, for optimal performance. Each heatpipe is secured to the outer surface of the copper mounting blocks with a copper clamp (Figure 1d) tightened via vented 8-32 screws, which provides high thermal transfer. The heatpipe feedthrough assemblies occupy the four 1.33" CF ports that surround the top 4.5" CF port, and transfer the heat from the electromagnet to the air-side watercooling system.

Finally, we use a water loop formed by a 200 W water chiller (Thermotek 255p) wired in series with square 40x40 mm consumer copper cooling blocks to dissipate the heat carried by the heatpipes. A notched aluminum plate acts as a heat exchanger between the heatpipe and the water block. This assembly is clamped together using aluminum plates, and secured to the chamber via an aluminum corner brace (Figure 1a and 1c).

The electromagnet assembly is designed to operate in the ultra-high vacuum (UHV) environment needed for atomic experiments, with vacuum pressures below $10^{-9}$ torr. This is achieved by using construction materials with low-outgassing characteristics. We use conductors, such as copper wire and OFE copper 101, and insulators, such as kapton and PEEK. The fasteners are vented to prevent virtual leaks. The construction and assembly uses simple methods to achieve thermal contact through direct mechanical contact. All materials are chosen to prevent excess background magnetism.

# Methods and Results

## Magnetometry

To characterize the magnetic field produced by the electromagnet, we perform a direct measurement with a Hall gaussmeter (F. W. Bell Model 5070). The probe consists of a mm-sized sensor attached to the end of a four inch long rectangular-profile arm. The one-axis vector probe measures the magnetic field component that is transverse to the arm length in the direction parallel to the narrow edge. The gaussmeter electronic offset was removed with a zero gauss chamber before each measurement.

We measured the spatial distribution of the magnetic field by moving the probe along the *x*, *y* and *z* axes using translation stages (Newport 9063-X-P operated manually). The motion of the stage is actuated with micrometer screws with 80 threads per inch.

Each geometrical axis was probed in separate measurements. Each measurement started by aligning the sensor to the center of the coil such that $x=y=z=0$, which we determine precisely by turning on the coil and zeroing all fields ($B_x$, $B_y$, $B_z$). We aligned the probe arm to be visually approximately along the axis of the measurement (*x* for example). To align the probe B-field sensitive axis with the experiment, we turn on the coil to a low current. For aligning the magnetically sensitive axis of the probe to the *z* axis of the coil, we offset the probe in the *z* direction and rotate the probe until the measured magnetic field value is maximized. For measuring the other transverse component of the field (either $B_x$ or $B_y$), the probe was rotated by 90 degrees from this axis with a rotation stage.

To obtain all three components of the magnetic field ($B_x$, $B_y$ and $B_z$) along all three electromagnet symmetry axes, we aligned the probe arm along two axes (*x* and *y*) and, in each case, scanned the position of the probe along all three axes (*x, y*, and *z*). For each scan, we measured both of the available transverse components of the magnetic field. This produces twelve measurements, out of which three are degenerate measurements that we use to check the accuracy of our procedure.

For each scan, the probe was translated to sequential positions along the measurement axis. For each position, approximately 20 magnetometer recordings (each retrieved every 0.5 s) were taken with field on (applied current $I$=20 A) and off ($I$=0 A), respectively. This procedure subtracts out drifts in the laboratory environmental magnetic field or in the electronic offset of the magnetometer. The data from each experimental configuration is averaged using standard gaussian statistics.

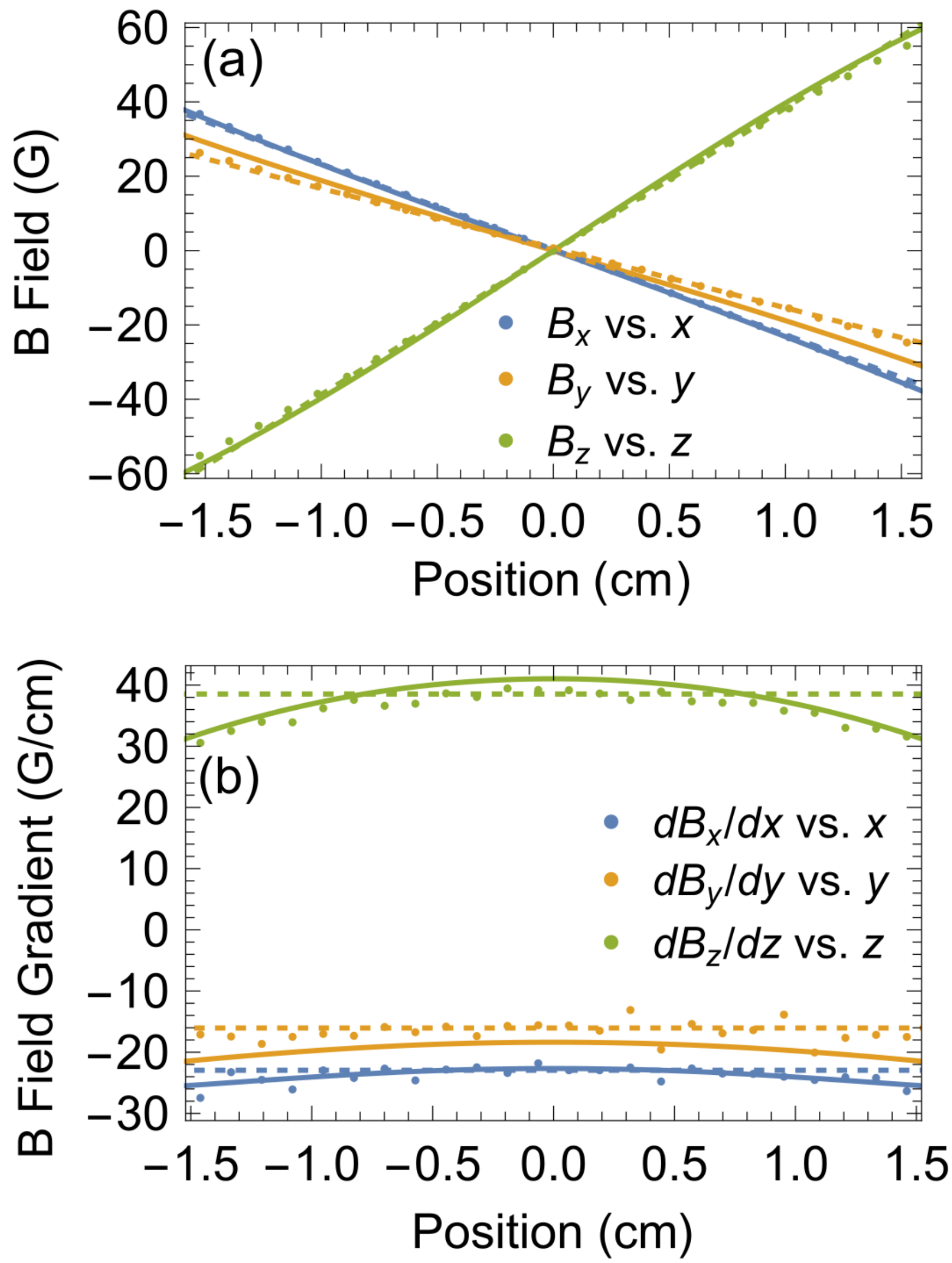

**Figure 2. Magnetic fields.** (a) Measured magnetic field components parallel with the three geometrical axes *x, y,* and *z* with the coils set to *I*= 20 A. (b) Magnetic field gradients are calculated from differences in the magnetic field between adjacent points. Solid lines show simulated magnetic fields and dashed lines show a least-squares linear fit to the six points of data near the origin. All measurement error bars are smaller than or at most comparable to the size of the data points.

The measured magnetic field components $B_x$, $B_y$ and $B_z$ as a function of position along their respective axis are shown in Figure 2a. The gradients shown in Figure 2b are extracted from the same data by taking differences between adjacent data points. Measurements of all of the other magnetic field components along all other geometrical axes can be found in the Appendix C. The data show an almost linear dependence of the field vs position, consistent with uniform gradients. The maximum magnetic field gradient of 38 G/cm is sufficient for levitating the Cs atoms as part of the evaporation scheme.[18] The gradient changes by less than 1% over the mm-sized atom cloud.

As shown in Figure 2, the measured magnetic fields are in good agreement with the prediction of a numerical simulation. The simulation uses an open-source Python package, Magpylib, to create the elliptical geometry of the coils and to calculate the magnetic fields and gradients. The simulation assumes stacked elliptical loops of current, which is a good approximation for the actual experimental geometry. Deviations between the measured magnetic field values and simulation can be explained by a combination of (1) imperfections in the manufacturing process, including machining and alignment of the mounts and wire wrapping procedure and (2) imperfections in magnetic field measurement i.e. from misalignment of probe and coil axes.

## Coil Switching Speed

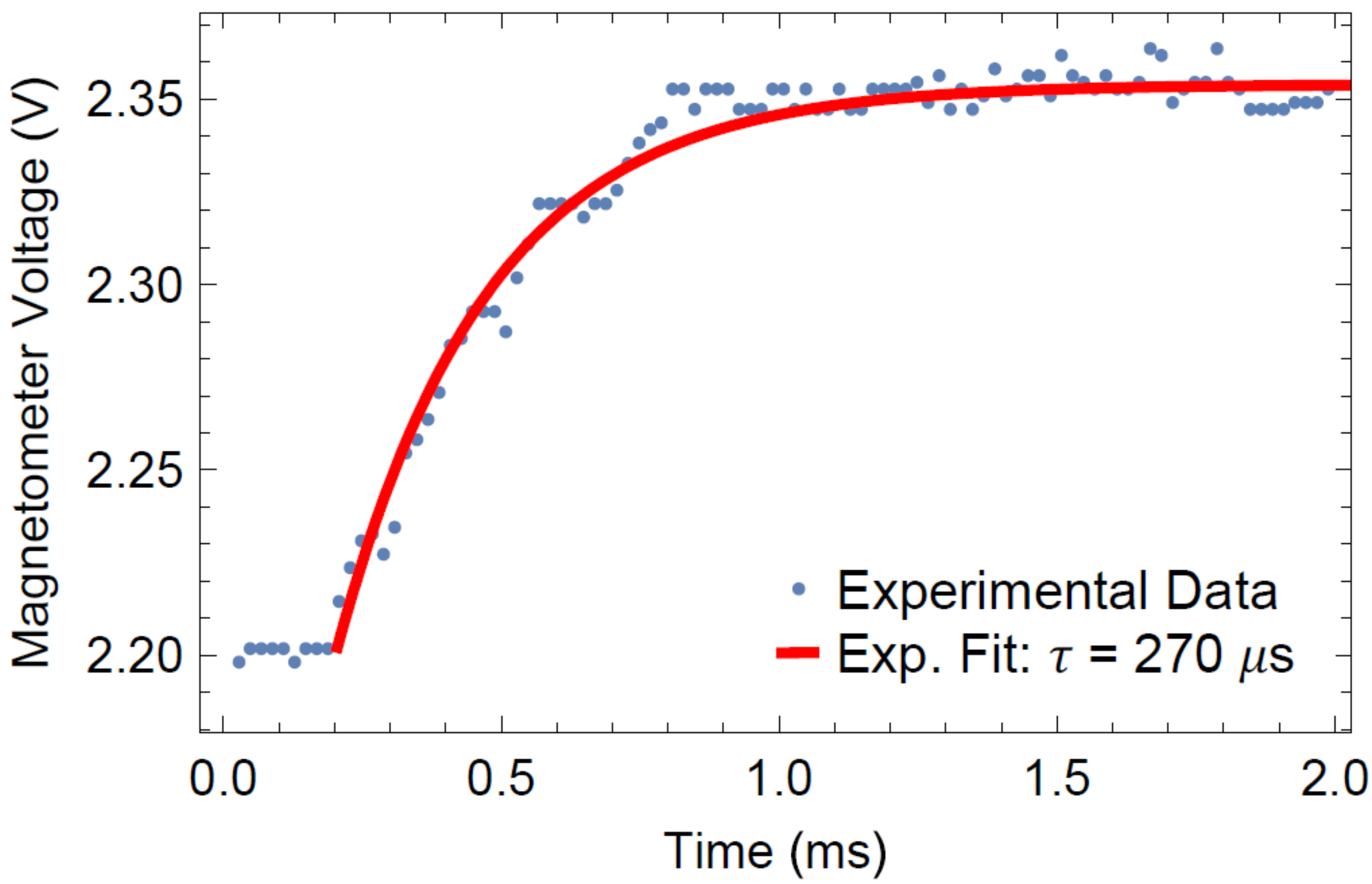

**Figure 3. Change in the measured magnetic field as the coils are switched on**. A least-squares fit to a decaying exponential gives an 1/*e* time of 270 us.

Transitions between several steps in the laser and evaporative cooling atom sample preparation sequence occur much more efficiently with the ability to switch on/off the magnetic field within the timescale of one ms. This is partially so that the atoms do not accelerate too much during the hand-off between the optical dipole and magnetic levitation steps.[18] For out-of-vacuum electromagnets, the switching time can be several to tens of ms,[2,21] limited by Eddy currents induced in the experiment vacuum chamber.

Fast switching is facilitated by placing the electromagnet inside the vacuum chamber in two ways: (a) induced Eddy currents are minimized, since the amount of electrically conductive metal near the electromagnet is significantly reduced and (b) the lower used power can be provided by simple current sources where feedback can be easily implemented. We break the primary Eddy current loops around the copper coil mounts using non-conductive PEEK spacers to split each coil into two separate segments.

We quickly switch the current applied to the coils using a power MOSFET (IXYS IXFN180N10 - switching time < 250 ns). A standard function generator provides a square pulse to trigger the electromagnet switch on/off. A magnetic sensor (Honeywell HMC2003) was used to measure the switching-on time of our electromagnet. The sensor was mounted to an aluminum post to avoid background magnetism and placed between the two coils of the apparatus.

The output of the sensor was captured on an oscilloscope (Rigol DS1104Z) and is shown in Figure 3. We use standard nonlinear least-squares to fit the data to a decaying exponential function and we find that the 1/*e* switching time equals 270 us. Full turn on to 99% of the final value occurs in less than 1 ms. We note that these measurements occur near the frequency bandwidth limit (~1 kHz) of the magnetic sensor (Honeywell HMC2003) and therefore might represent upper limits. Nevertheless, the measured switching times are faster than similar uncompensated out-of-vacuum setups. They are sufficient for the quantum control steps used in our experiment for atom cooling and sample preparation.

## Thermal and Vacuum Testing

The thermal performance of the electromagnet is tightly tied to the lowest achievable pressure of the vacuum system. As the in-vacuum coils heat up during operation, outgassing of the hot surfaces become the primary source of background gas pressure.

### Thermal imaging in air

We perform a preliminary evaluation of the cooling system in air. We characterize the heat-flow in diagnostic measurements where we record thermal images (Figure 4) using a commercial thermal camera (FLIR T560). This test was performed with the same electromagnet coils and mounts as the in-vacuum setup described above, but with a different set of four commercial heatpipes that did not have vacuum feedthroughs attached. The heatpipes have identical dimensions to the ones used above, are water-filled and have a heat capacity of 35 W each. In order to perform accurate thermal imaging, the apparatus was covered with masking tape to provide a non-reflective surface to image. This test was performed in air before vacuum cleaning. The coils were driven with 20 A each.

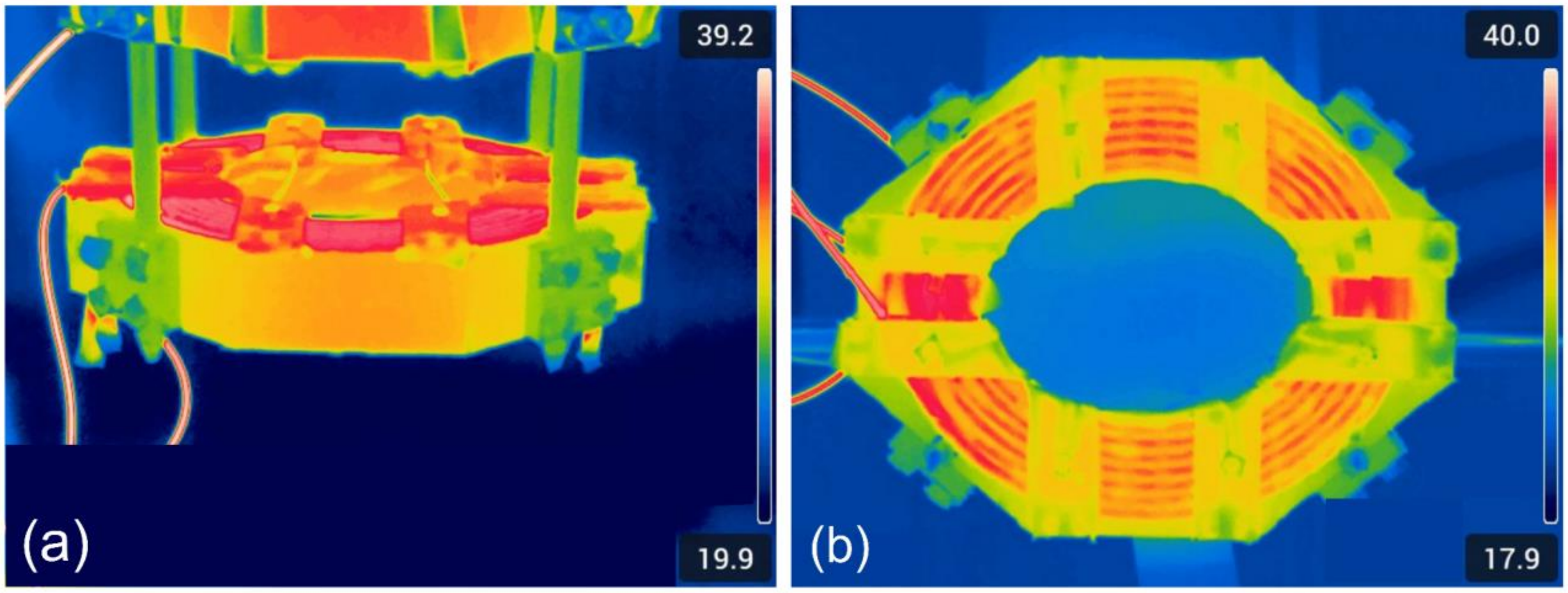

**Figure 4. Thermal images of the in-vacuum electromagnet apparatus.** Side view (a) and top view (b). The temperature scales were automatically generated by the commercial thermal camera (FLIR T560). The coils are driven at $I$=20 A in air.

The thermal images let us identify the primary heat flow paths and thermal resistances of various components from the measured temperature gradients. The largest gradient is between the kapton coil wire and copper mount (~10° C), consistent with the relatively large thermal resistance of the wire kapton insulation (0.2 W/(m·K) ) compared to the bulk copper coil (400 W/(m·K) ). A lower gradient of 2-3° C is observed between the center and the boundary of the copper coils. The gradients between the copper coils and heatpipes (3-4° C) correspond to the pressure-based thermal interface provided by the clamps. There are minimal thermal gradients (<1-2° C) along the heatpipes, consistent with very high thermal conductivity. The aluminum connections between heatpipes and the copper block that is part of the water cooling loop show thermal gradients of 4-5° C. Improved clamping of the heatpipes on the air side could reduce these gradients in the future.

### Thermal tests inside the vacuum chamber

Controlling environmental heat dissipation through convection is very hard to manage in air, so a test with the coils inside the vacuum chamber is necessary to fully evaluate thermal performance. The in-vacuum test uses the custom-made heatpipe vacuum feedthroughs (Noren Thermal) in the complete experiment configuration, as described in the construction section and shown in Figure 1. Because of a mechanical failure of one of the four heatpipes that caused a vacuum leak at the $10^{-6}$ torr·L/s level, only three heatpipes were used in the test below.

We determine the temperature in two ways. First, the core wire temperature is calculated from the relative change in resistance of each coil with temperature. The initial resistance was calibrated before each measurement by measuring the electromagnet wire resistance after it had thermalized to the known room temperature. Second, the temperature of the copper coil mount is read at five points using PT100 thermistors (Allectra 343-PT100-C1) secured to the mount via BeCu clamps. The in-air water loop was set to 12° C for all tests.

We measure the vacuum pressure of our system using an extended range ion gauge (Agilent UHV-24p). The background pressure with the electromagnet turned off was below $8 \cdot 10^{-11}$ torr, provided by a combo ion-getter pump system (SAES NEXTorr Z500). This pressure was achieved after a week-long bake out of the vacuum chamber at 110° C.

We operated the electromagnet with various heat loads and measured the increase in electromagnet temperature and associated increase in vacuum pressure from outgassing of absorbed gas in the coil and chamber materials. We observe strong exponential scaling of vacuum pressure with coil temperature (Figure 5), consistent with literature.[26] In our experiments, we expect that the atoms will spend a few seconds in the vacuum chamber during atom sample separation, which sets a maximum vacuum pressure of $10^{-9}$ torr for minimal atom loss.[27] This corresponds to electromagnet temperatures during operation that are lower than $T_{max}$ = 60° C.

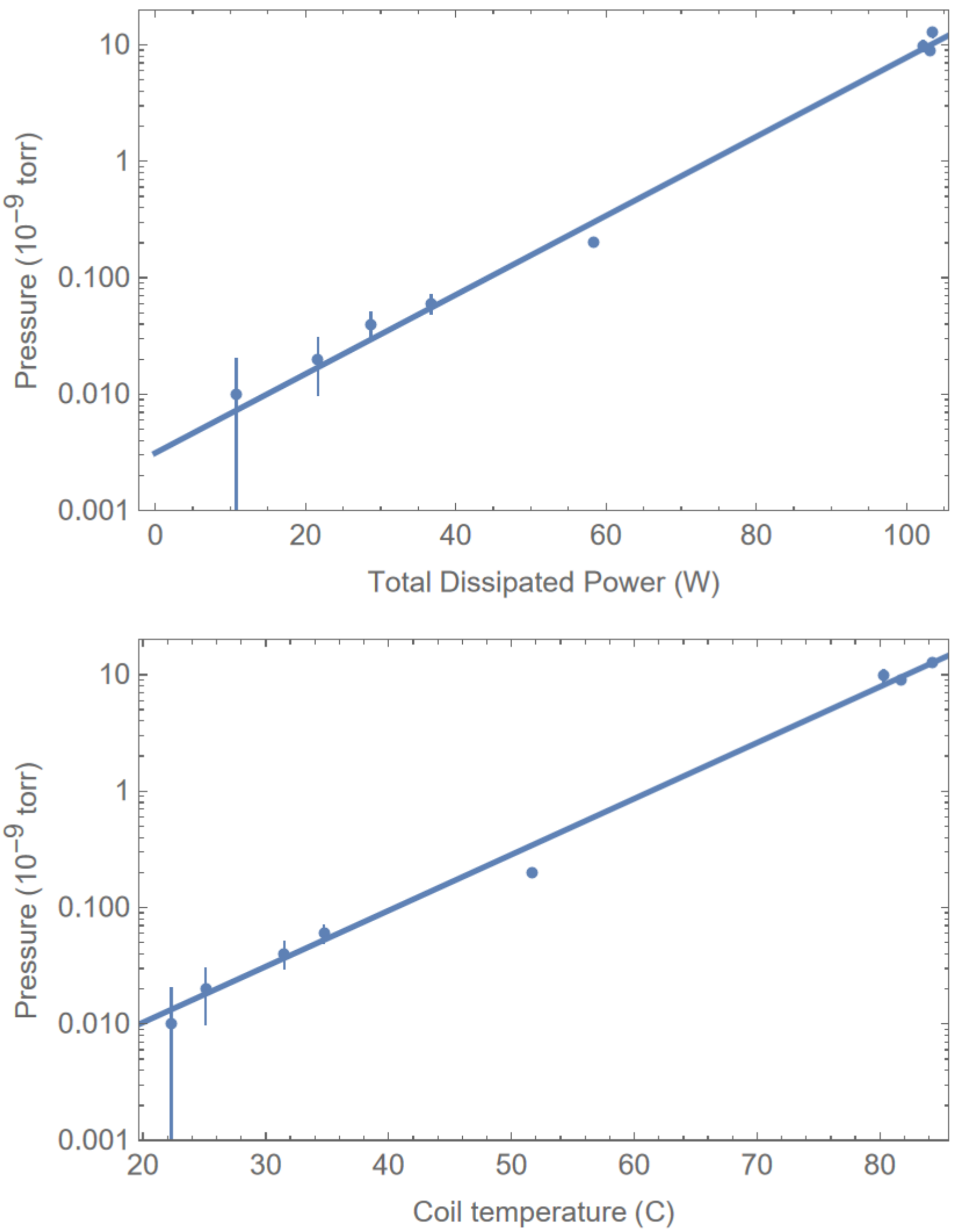

**Figure 5.** Measured dependence of the pressure inside our vacuum chamber as a function of the dissipated power and temperature of the electromagnet. The solid line corresponds to least square fits. We observe exponential increase in pressure due to outgassing of the electromagnet and vacuum chamber surfaces.

We first test the electromagnet thermal performance with currents necessary to create the magnetic field gradients (15-20 G/cm) for the magneto-optical trap (MOT). At 7 A, the electromagnet dissipates 22 W, which heats the wire core to a temperature of 25° C that stabilizes after 2-3 hours. The corresponding increase in vacuum pressure of $2 \cdot 10^{-11}$ torr is compatible with minimal atom loss due to background gas collision.

Producing the levitating magnetic field gradient (31 G/cm) necessary in the hybrid evaporation cooling scheme requires running the electromagnet at 16 A with ~50% duty cycle, which accounts for the MOT loading and other experimental cooling and detection steps. We simulate these conditions by operating the electromagnet at 11 A, which dissipates 58 W. With these settings, we observed that the wire core temperature settles after 4-5 hours to an average of 52° C and vacuum pressure settles to $2 \cdot 10^{-10}$ torr. These measured pressures are low enough so that minimal loss in atom numbers due to background gas collisions are expected during atom sample preparation. Throughout testing, we observed that the electromagnet heating rate is much higher than expected given typical heatpipe thermal conductivities, particularly when running the coils with heat loads above 60 W. This is due to non-uniform thermal performance of the heatpipes. A detailed thermal model of the system is presented in Appendix D and suggests that thermal dissipation up to 200 W is possible with four high-conductivity heatpipes. Better thermal anchoring might allow heat dissipation of up to 300 W.

## Discussion and Conclusion

Further increases in maximum heat load may be achieved by upgrading the thermal interfaces using more advanced manufacturing processes such as brazing or with high performance materials, such as electrical insulators with good thermal conductivity such as aluminum oxide (alumina) or aluminum nitride.[4]

In conclusion, we have described the design and implementation of an in-vacuum electromagnet for quantum science experiments. Our design produces magnetic fields and gradients as high as 130 G and 38 G/cm, respectively, sufficient for MOT and magnetic levitation applications. Heatpipe assisted cooling dissipates 50 W of thermal power with minimal increase in temperature in a passive, compact, low-vibration apparatus. The resulting outgassing rate allows operation at UHV pressures, below $10^{-9}$ torr. The coil can be switched in less than 1 ms. The technology of heatpipe cooled electromagnets is applicable to various quantum science experiments that use atomic and molecular MOTs, magnetic trapping or evaporative cooling.[1–6]

## Acknowledgements

We thank B. Anderson, C. Condos, J. Jones, H. Muller, S. Nobel, S. Pau, R.Schickling, D. Wilson and J. Wu for sharing equipment, technical assistance and valuable discussions.

# Author Declarations

## Conflict of Interest

The authors have no conflicts to disclose.

## Author Contributions

**Kenneth Nakasone**: Methodology (equal), Investigation (equal), Validation (lead), Visualization (supporting), Writing - Original Draft (equal)

**Paola Luna**: Investigation (supporting), Software (equal), Visualization (supporting), Writing - Original Draft (supporting)

**Andrei Zhukov**: Investigation (supporting), Writing - Review & Editing (equal)

**Matthew Tao**: Conceptualization (equal), Writing - Review & Editing (equal)

**Garrett Louie**: Conceptualization (supporting), Writing - Review & Editing (equal)

**Cristian Panda**: Conceptualization (equal), Methodology (equal), Investigation (equal), Formal Analysis (lead), Visualization (lead), Writing - Original Draft (equal), Project Administration (lead)

## Data Availability

The data that supports the findings of this study are available from the corresponding authors upon reasonable request.

# Appendix A: Optimization of coil separation

The electromagnet geometry was optimized by measuring the magnetic field gradient $dB_z/dz$ with $I$=10 A while varying the distance between the inner surfaces of the two coil mounts, d (Figure 6). The largest gradient is obtained for $d$= 35 mm but we chose a $d$= 46 mm separation for our experiment to maximize optical access between the coils. All other measurements shown here correspond to this separation.

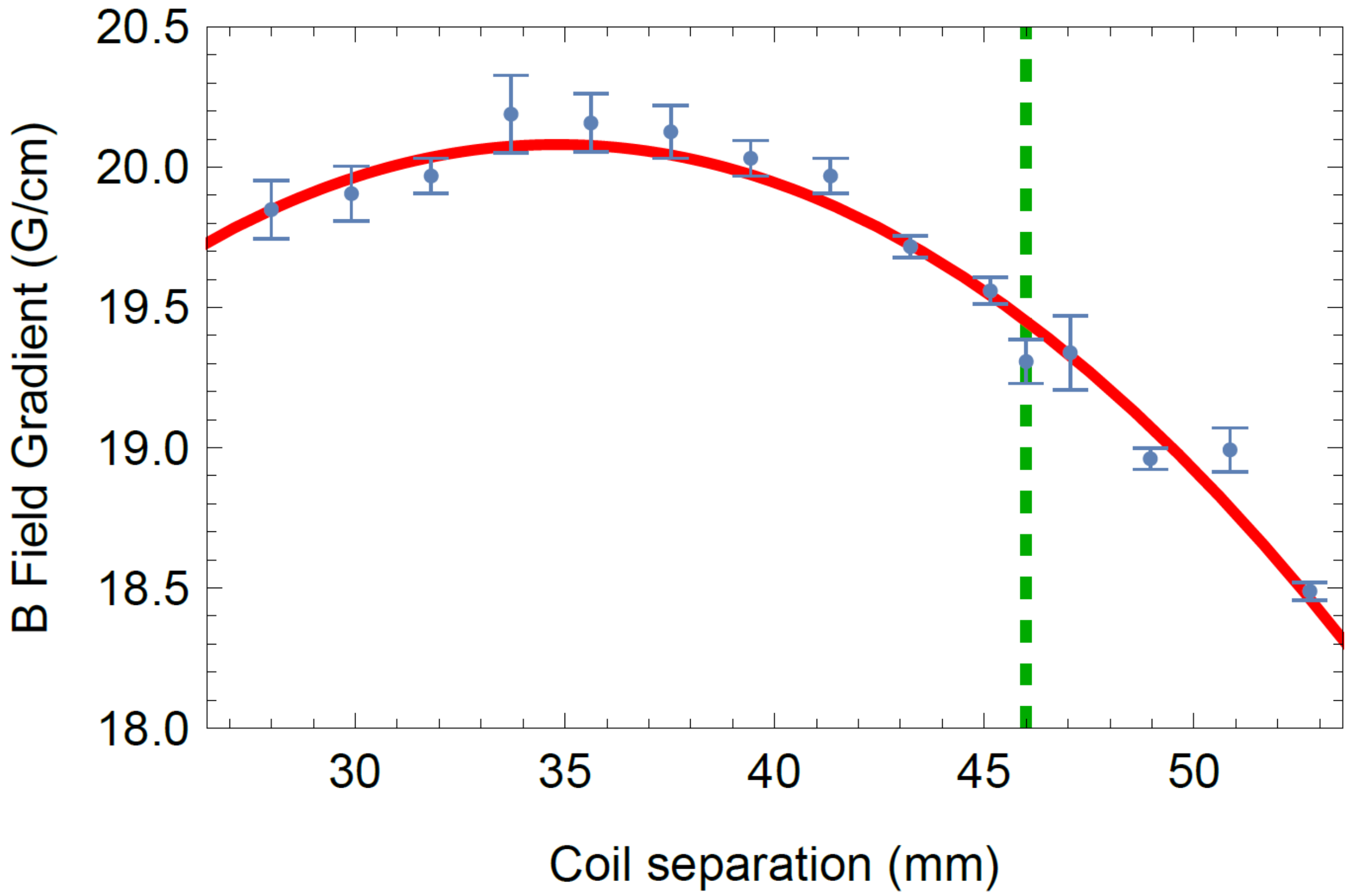

**Figure 6. Coil separation optimization.** Measured values of the magnetic field gradient for $I$= 10 A as a function of separation distance between the two electromagnet coils and quadratic fit (red line). All error bars represent 1σ of a normal distribution (68% confidence interval). The vertical green dashed line shows the separation we chose for our electromagnet geometry, which maximizes optical access over field magnitude.

# Appendix B: Two-Piece Arm Assembly

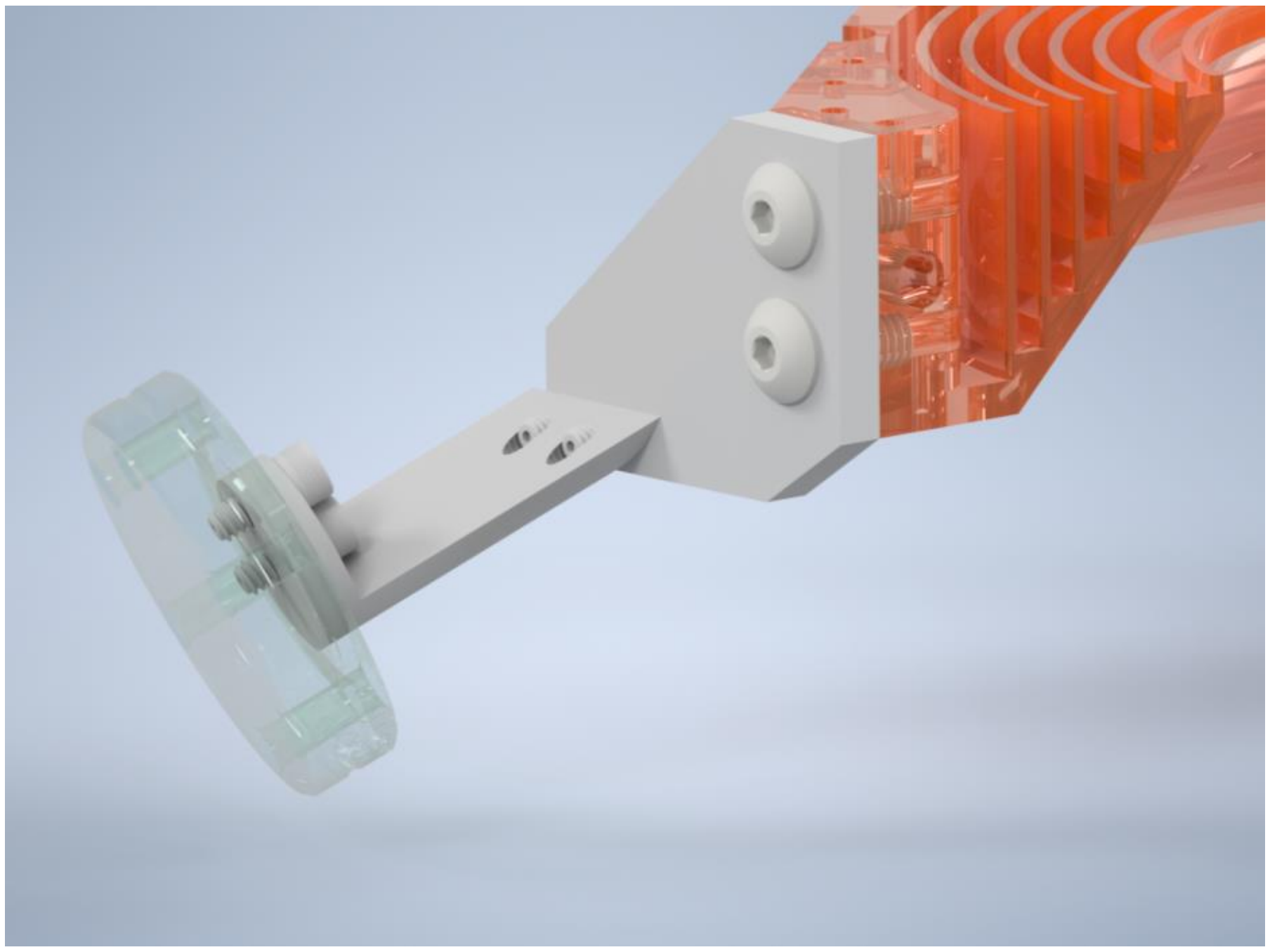

**Figure 7.** Render of the two-piece arm assembly for securing copper mounting block to the vacuum chamber. The assembly consists of two pieces: a half-cylinder post (middle left) and a triangular wing (right). The half-cylinder post is attached to a 1.33” conflat blank (left, transparent), securing the copper mounting blocks to the vacuum chamber. To adjust the centering of the copper mounting blocks, a shim is inserted between the triangular wing and the copper block.

The copper mounting blocks are secured to the Kimball vacuum chamber using the two-piece arm assemblies depicted in Figure 7, consisting of a half-cylinder post and triangular wing. A triangular wing (center right) attaches to #8-32 tapped holes on the outer surface of the copper mounting block (right, transparent orange). The half-cylinder post (center left) attaches to two #4-40 holes in a 1.33” CF blank (left, transparent), thereby securing the copper mounting block to the vacuum chamber. The half-cylinder and wing are secured to each other using #4-40 screws.

The blanks are installed into the 1.33” CF ports that surround the y-axis 4.5” CF ports of the vacuum chamber. The copper mounting blocks are centered in the vacuum chamber by inserting 316 stainless steel shims between the triangular wing and copper mounting block. All screws depicted are vented for vacuum compatibility.

# Appendix C: All Magnetic Field Measurements

In addition to the fields described in the section titled “Magnetometry” above, we also measured all 9 possible magnetic field components along the three geometrical axes x, y, and z using the same methods described above (Figure 8). Small offsets in the fields perpendicular to the quadrupole axes are due to misalignment between the measurement probe and coil axes.

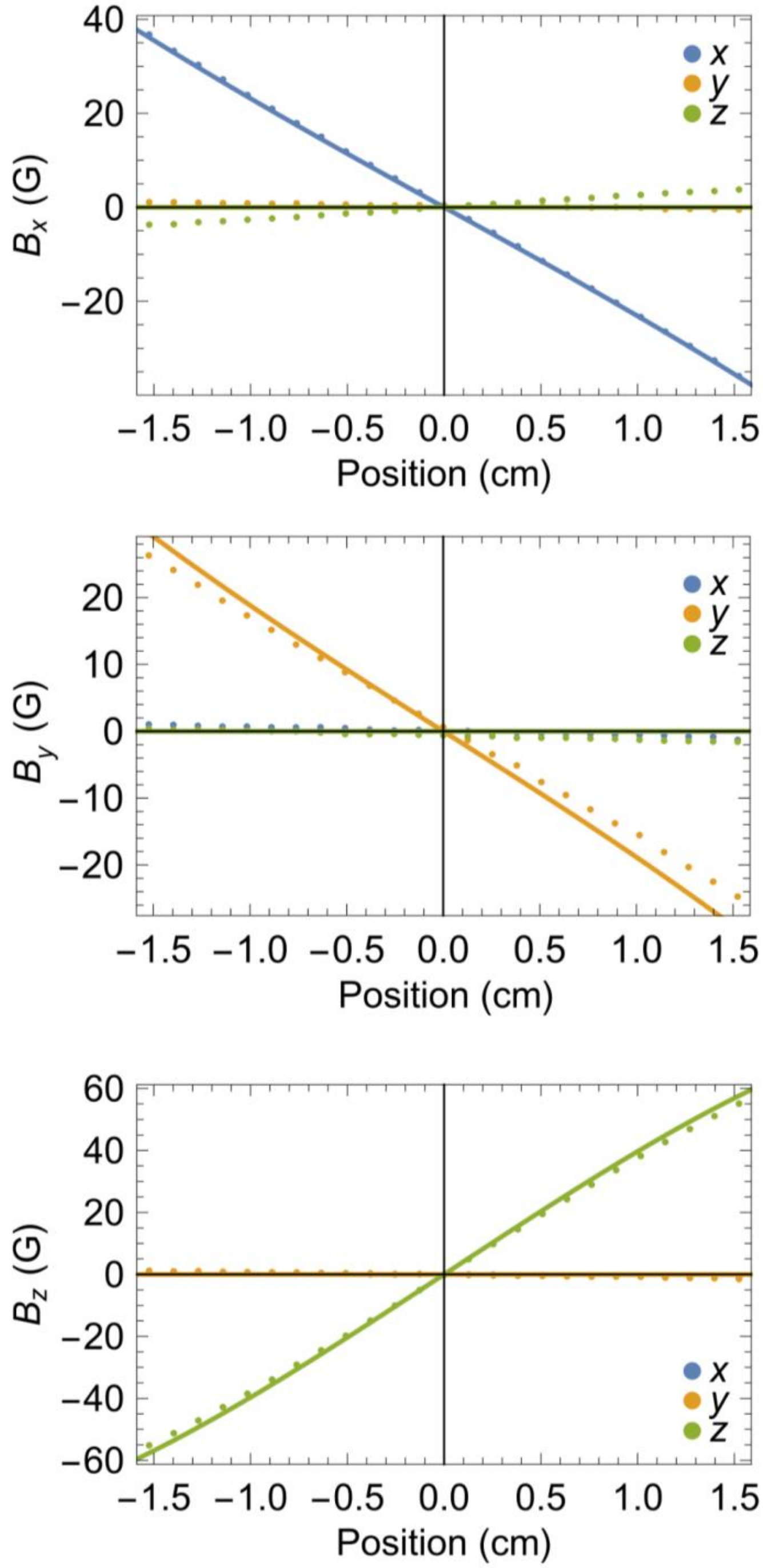

**Figure 8.** Measurement of all 9 possible magnetic field components along the three geometrical axes *x*, *y*, and *z* with the coils at $I$= 20 A. Magnetic field simulations are shown as solid lines.

We also switched the coils to a Helmholtz configuration, where current is flowing in the same direction for both coils. We measured the largest magnetic field component, $B_z$, along all three axes and observed less than 1% variations in the magnitude of the field over the central 1 mm region corresponding to the location of the atom sample. The data (Figure 9) is consistent qualitatively and the measured magnetic field magnitude is within 3% of the simulations.

The measured curvature is consistent with a few factors: (1) the coils are elliptical rather than the typical circular shape, (2) the distance between the coils is larger than the mean radius of the coils and (3) the windings are spaced by a mm and the thickness of the wires is a significant fraction of the coil dimensions. Despite all these, the magnetic field value varies by less than 1 part in 1,000 over the size of the atom sample (+-1 mm).

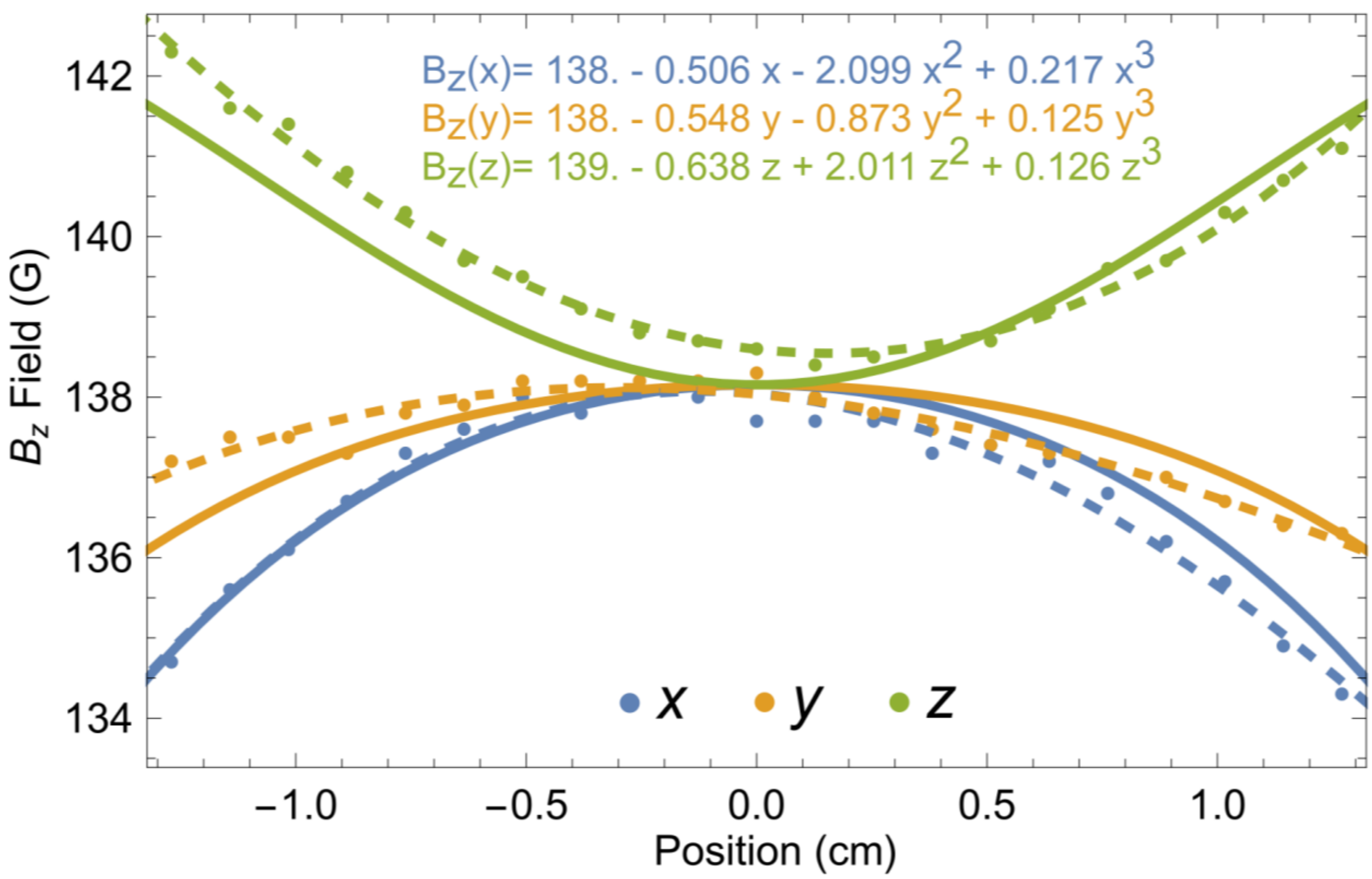

**Figure 9.** Measured magnetic field z components with electromagnet operating in the Helmholz configuration, where both coils carry equal currents in the same direction. Solid lines show the simulated magnetic field with magnitudes scaled by 0.977. Dashed lines show least squares fits to third order polynomials. The equations of these fits are displayed at the top of the plot.

# Appendix D: Thermal model, limitations and possible upgrades

We improved our understanding of the electromagnet thermal limitations by modeling the thermal flow in our system (Figure 10), without needing to disassemble the electromagnet.

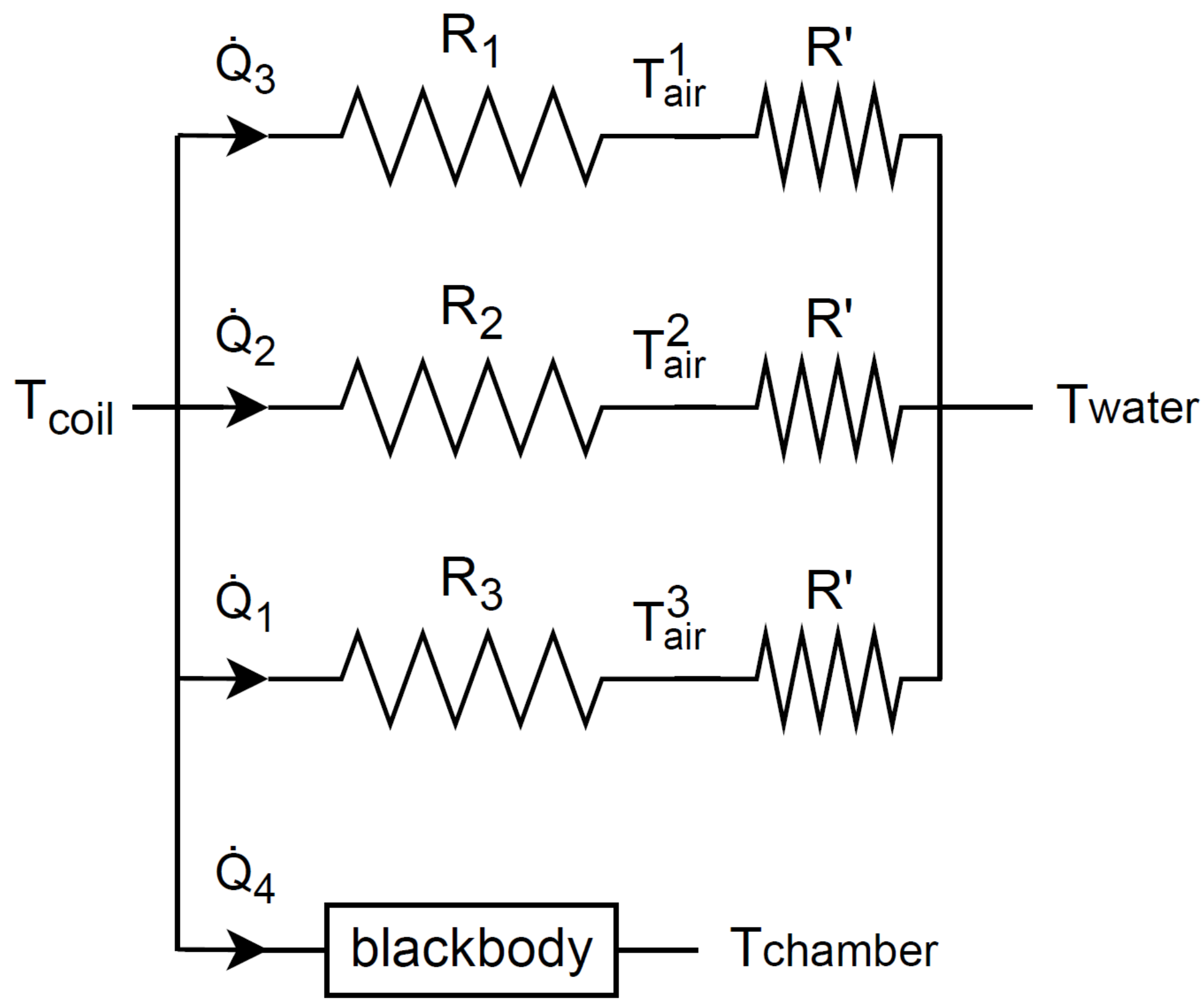

**Figure 10. Thermal model diagram.** Heat flow from the electromagnet ($T_{coil}$) follows three paths along heapipes (resistances $R_1$, $R_2, R_3$) to dissipate heat into the water loop ($T_{water}$). At large temperature differences ($T_{coil}$ - $T_{chamber}$ > 40° C), a fourth thermal flow path due to blackbody radiation to the vacuum chamber ($T_{chamber}$) becomes significant.

In analogy with an electrical circuit, the three heatpipes act as a parallel thermal resistances ($R_1$, $R_2, R_3$) that carry away heat ( $\dot{Q}_1$, $\dot{Q}_2$, $\dot{Q}_3$) from the electromagnet (temperature $T_{coil}$) to the air-side heatpipe end ( $T_{air}^1$, $T_{air}^2$, $T_{air}^3$). Each heatpipe is in series with the thermal resistance ($R'$) of the interface between the heatpipe and the water cooling block (temperature $T_{water}$). We assume all $R'$ values are approximately equal. Independent measurements support this assumption at the 10- 20% level.

At higher coil temperatures, blackbody radiation becomes a significant heat transfer mechanism. We model this as a fourth thermal dissipation path $\dot{Q}_4$ from the coil ($T_{coil}$) to the vacuum chamber ($T_{chamber}$) using the Stefan-Bolzman law: $\dot{Q}_4 = \sigma(T_{coil}^4 - T_{chamber}^4)$, where $\sigma = k \cdot A = 3.4 \times 10^{-7}$ W/K$^4$ is the Stefan-Boltzmann constant $k = 5.67 \times 10^{-8}$ W/(m$^2$·K$^4$) multiplied by the area of the electromagnet $A = 0.06$ m$^2$. For our target $T_{max}$= 60° C and when $T_{chamber}$=

20° C, blackbody radiation transfers $\dot{Q}_4$= 17 W, a significant fraction of the total electromagnet power.

The total electromagnet dissipated heat is then $\dot{Q} = \dot{Q}_1 + \dot{Q}_2 + \dot{Q}_3 + \dot{Q}_4$. In addition to the measured quantities already described above, we measure $T_{air}^1$, $T_{air}^2$, $T_{air}^3$, $T_{water}$ and $T_{chamber}$ with standard K-type thermocouples. We use simple circuit analysis to extract the thermal resistances of the three heatpipes $R_1$, $R_2, R_3$.

We performed measurements at electromagnet power loads ($\dot{Q}$) that varied from a few to 100 W. For all heat loads, we observe that $T_{air}^1$ is consistently much higher than $T_{air}^2$ and $T_{air}^3$, both of which remain close to the temperature of the water reservoir. This suggests that $R_1$ is much smaller than $R_2$ and $R_3$. For relatively large thermal loads ($\dot{Q}$> 50 W), we use our model to extract thermal resistances of $\{R_1, R_2, R_3\} = \{0.5,3,12\}$ K/W, which correspond to thermal conductivities $\{K_1, K_2, K_3\} = \{14, 2, 0.6\}$ kW/(m·K).

We find that $R_1$ is consistent with typical heatpipe thermal conductivity values,[24] while the other two heatpipes resistivities, $R_2$ and $R_3$ are very high. The reasons for these differences in performance are unknown, but we speculate they are due to variations in the manufacturing process.

The low thermal heatpipe conductivities are the primary limiting factors to the thermal performance of our system. Replacing the underforming heatpipes should drastically reduce heating. We estimate the thermal performance of the electromagnet with four heatpipes with resistance that equals the high-performing heatpipe, $R_1 = R_2 = R_3 = R_4 = 0.5$ K/W. Maintaining the temperature of the coil to below $T_{max}$ = 60° C is possible while dissipating 200 W, about 3.5 times more power than in our current setup. This power dissipation would be sufficient to generate gradients of 50 G/cm or field offsets above 150 G with 100% duty cycle.

Further improvements in thermal performance may be achieved by increasing thermal conductivity in the air side connection from heatpipes to the water block ($R'$) with a simple copper connection water block or by brazing the heatpipes to the water block. If $R'$ can be lowered two-fold, heat dissipation of 300 W is possible, sufficient for generating field gradients of 60 G/cm or field offsets above 200 G.